\documentclass[aps, pra, twocolumn]{revtex4-2}

\usepackage{graphicx}
\usepackage{amsmath, braket}
\usepackage{hyperref}

\begin{document}
    \title{Quantum interference induced magnon blockade and antibunching in a hybrid quantum system}

    \author{Pooja Kumari Gupta}
        \email{pooja.kumari@iitg.ac.in}
    \author{Sampreet Kalita}
        \email{sampreet@iitg.ac.in}
    \author{Amarendra K. Sarma}
        \email{aksarma@iitg.ac.in}

    \affiliation{Department of Physics, Indian Institute of Technology Guwahati, Guwahati-781039, Assam, India}

    \begin{abstract}
        In this work, we study the phenomena of quantum interference assisted magnon blockade and magnon antibunching in a weakly interacting hybrid ferromagnet-superconductor system.
        The magnon excitations in two yttrium iron garnet spheres are indirectly coupled to a superconducting qubit through microwave cavity modes of two mutually perpendicular cavities.
        We find that when one of the magnon mode is driven by a weak optical field, the destructive interference between more than two distinct transition pathways restricts simultaneous excitation of two magnons.
        We analyze the magnon correlations in the driven magnon mode for the case of zero detunings as well as finite detunings of the magnon modes and the qubit.
        We show that the magnon antibunching can be tuned by changing the magnon-qubit coupling strength ratio and the driving detuning.
        Our work proposes a possible scheme which have significant role in the construction of single magnon generating devices.
    \end{abstract}

    \maketitle

    \section{Introduction}
        \label{sec:intro}
        In the past decade, quantum systems based on light-matter interaction have seen several applications in quantum information processing, quantum sensing and quantum communication \cite{Nature.416.238, RevModPhys.89.035002,NaturePhoton.1.165}.
        To achieve greater control over the quantum phenomena observed in such systems and enhance their scalability, hybrid cavity quantum mechanical systems that couple optical, microwave, and mechanical degrees of freedom have also been studied extensively \cite{RevModPhys.85.623, PNAS.112.3866}.
        Recently, such hybrid systems have also introduced interactions with magnons \textemdash{} quanta of collective excitations in magnetic materials \cite{PhysRevApplied.2.054002, PhysRevLett.113.156401, PhysRevLett.113.083603}.
        This emerging area of cavity quantum systems, commonly known as cavity optomagnonics \cite{PhysRevA.94.033821}, explores the coherent interaction of magnons with different degrees of freedom, such as optical photons, microwave photons, phonons, and superconducting qubits inside a cavity\cite{PhysRevLett.116.223601, AAAS.349.405, APEX.12.070101, PhysRevLett.121.203601}.
        The various kind of interactions like magneto-optical \cite{PhysRevB.93.144420, PhysRevLett.121.087205}, magnetic dipole\cite{ApplPhysLett.108.062402, CRPhys.17.729}, and magnetostrictive interaction \cite{PhysRevA.99.021801, SciAdv.2.e1501286, NewJPhys.21.085001}provide a versatile platform to develop innovative technologies like a microwave to optical transducers for quantum communication \cite{Optica.7.1291, PhysRevB.93.174427, NewJPhys.23.043041}, single magnon-generating sources for quantum sensing \cite{SciAdv.3.e1603150, Science.367.425} and quantum information processing \cite{ApplPhysLett.122.084001, JApplPhys.128.130902}, quantum cryptography \cite{PhyRevB.105.094422} and quantum simulators \cite{arXiv.1903.12498}.
        Magnons' long lifetime and high tunability also make them a promising candidate for information carriers and in combining known quantum platforms \cite{PhysRevB.100.134402, PhysRevResearch.1.032023}.
        Theoretical studies have also explored phenomena like magnon-induced transparency \cite{OptExpress.26.20248, PhysRevA.102.033721}, nonclassical states of magnons \cite{PhysRevLett.127.087203, PhysRevA.100.013810}, magnon dark modes \cite{JApplPhys.126.173902, PhysRevB.99.094407}, slow light \cite{IEEEAccess.7.57047, OptExpress.27.5544}, magnon polariton bistability \cite{PhysRevLett.120.057202} and optical cooling of magnons \cite{FundamRes.3.3}.
        Experimental observation of the coherent interaction between magnon and qubit through virtual photon excitation of a microwave cavity mode has also been reported \cite{Science.349.405}.
        A promising ferromagnetic insulator used for this purpose is yttrium iron garnet (YIG), which features high spin density ($2.1 \times 10^{22} \mu_{B} \textrm{cm}^{-3}$), low damping rate and low absorption coefficient for electromagnetic radiation \cite{PhysRep.229.81}.
        YIG spheres are commercially available and are widely used for cavity optomagnonical experiments \cite{JPhysD.43.264002}.

        In the context of single magnon generation, it has been theoretically proposed that the nonlinear magnon-superconducting qubit interaction prevents the simultaneous excitations of more than one magnon inside the cavity, which leads to magnon antibunching and the phenomenon of magnon blockade \cite{JOSAB.38.876}.
        Magnon antibunching is a purely quantum effect and provides an ordered magnon distribution in space, which can be used to realize single magnon sources.
        Depending on the strength of magnon-qubit interaction, such an antibunching can result from two distinct types of physical mechanisms.
        For a strongly interacting nonlinear system, the quantum anharmonicity in the energy eigenspectrum leads to antibunching through conventional magnon blockade (CMB) \cite{JOSAB.38.876}.
        On the other hand, in a weakly-interacting nonlinear quantum system, unconventional magnon blockade (UMB) is observed, which arises due to quantum interference between two or more distinct quantum transition pathways \cite{PhysRevA.101.042331}.
        Magnon blockade was first theoretically proposed in Ref. \cite{PhysRevB.100.134421}, where they illustrated the dependence of blockade on the strength of the interaction.
        Following this, magnon blockade has been studied in various systems such as parity time-symmetric-like magnomechanical system \cite{AnnPhys.532.2000028}, ferromagnet-superconductor systems with two qubits \cite{PhysRevB.104.224434}, and other magnon based hybrid systems \cite{ApplPhysLett.121.122403, PhysRevA.103.052411}.
        
        In this work, we theoretically explore quantum interference-assisted magnon blockade and its antibunching characteristics in a hybrid ferromagnet-superconductor system containing two YIG spheres and a qubit.
        The magnon modes generated in the two YIG spheres interact individually with the qubit through virtual photon excitations, and  do not interact among themselves.
        Only one of the two magnon modes is driven by an external optical field.
        For a weak drive and weakly interacting magnons and qubit modes, we observe distinct two-magnon excitation pathways.
        These paths interfere destructively to produce magnon blockade.
        We show that we can tune the magnon antibunching by varying the parameters of the system.
        We further determine the optimal conditions for magnon blockade and discuss the case of magnon antibunching for (i) zero detunings and (ii) finite detunings of the magnon modes and the qubit.
        For either case, we observe that the antibunching is maximum when the coupling strength between the undriven magnon mode and the qubit is much less than that between the driven magnon mode and the qubit.
        The dependence of antibunching on the magnons-qubit coupling strength ratio is found to be non-monotonic.
        However, we can further control the degree of magnon antibunching for a fixed ratio of coupling strengths by varying the driven magnon detuning.
        Compared to previous proposed works on magnon blockade \cite{ApplPhysLett.121.122403}, our scheme does not require both the magnons to be driven optically.
        This substantially reduces the complications in experimental realization.

        Our work is structured as follows.
        In Sec. \ref{sec:model}, we introduce the model of our system and derive dynamic equations using the effective Hamiltonian.
        In Sec. \ref{sec:model_analytical}, the approximate analytical calculation and optimal equation for magnon blockade are discussed.
        The method for exact numerical calculation is presented in Sec. \ref{sec:model_numerical}.
        We then discuss our results for the two cases of antibunching in Sec. \ref{sec:results_zero} and Sec. \ref{sec:results_finite} and finally we study the effect of thermal magnons on the magnon blockade.
        The complete work is summarized in Sec. \ref{sec:conclusion}.

    \section{Model}
        \label{sec:model}
        \begin{figure}[ht]
            \centering
            \includegraphics[width=0.48\textwidth]{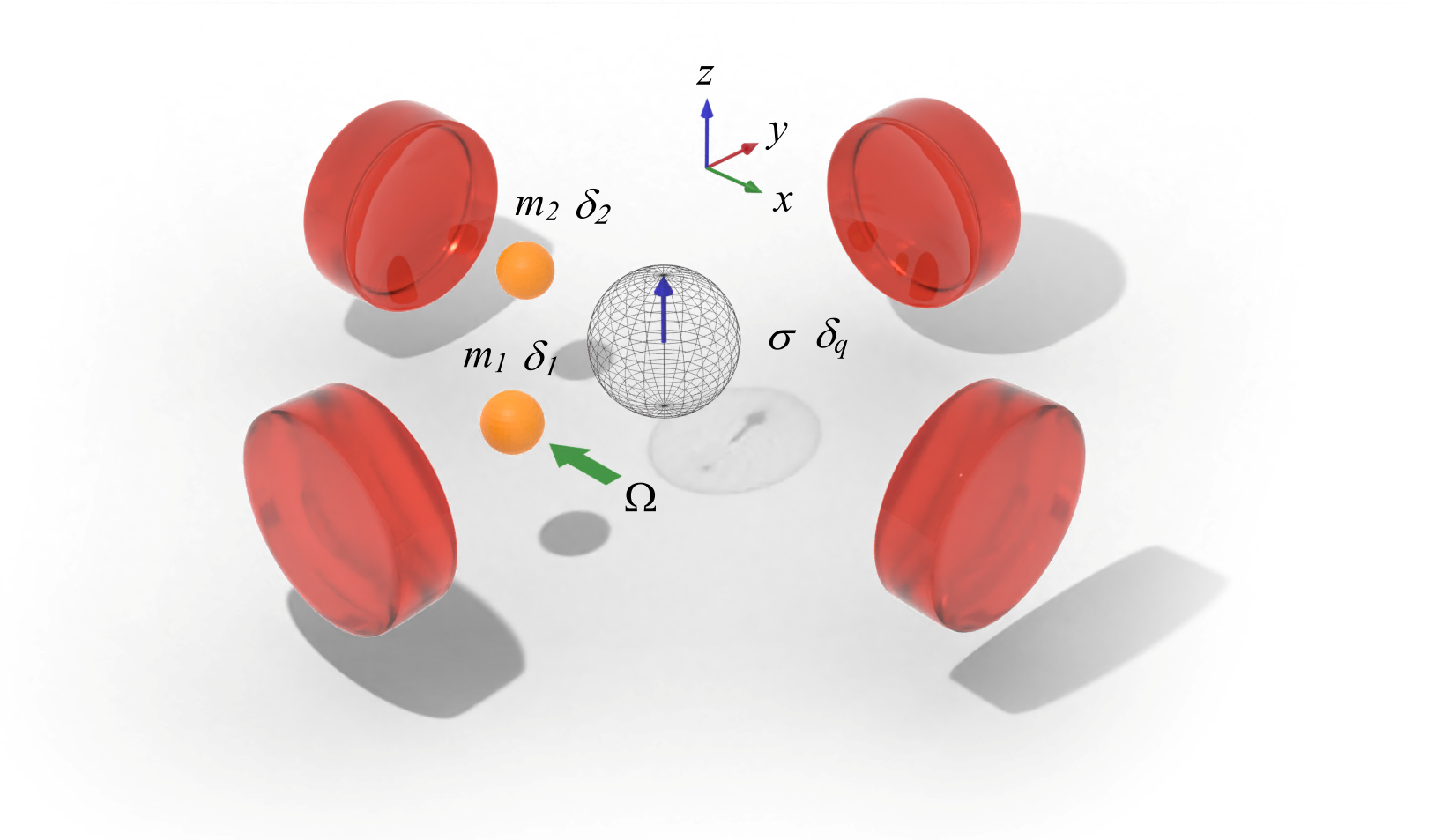}
            \caption{
                Schematic diagram of the hybrid system.
                Two YIG spheres inside the two cavities interact indirectly with the qubit at the center.
            }
            \label{fig:model}
        \end{figure}

        We consider a ferromagnet-superconductor hybrid quantum system containing a superconducting qubit and two YIG spheres inside two mutually perpendicular microwave cavities, as drawn schematically in Fig. \ref{fig:model}.
        The superconducting qubit is placed at the location of the maximum electric field of the two microwave cavities, and the two YIG spheres are placed where the magnetic field is maximum.
        
        A bias magnetic field $\mathcal{H}_{0}$ is applied in the z direction to magnetize the YIG spheres uniformly, and two local magnetic fields $\mathcal{H}_{1}$ and $\mathcal{H}_{2}$ are applied additionally to tune the frequencies of magnon mode $\hat{m}_{1}$ and $\hat{m}_{2}$ separately. The magnon modes in the two YIG spheres are indirectly coupled to the superconducting qubit via virtual photon excitations of the two cavity modes.
        One of the YIG spheres is driven by an optical field of frequency $\omega_{d}$ and strength $\Omega$.
        Its magnon mode $\hat{m}_{1}$ with frequency $\omega_{1}$ couples to the superconducting qubit with coupling strength $g_{1}$ whereas the other magnon mode $\hat{m}_{2}$ with frequency $\omega_{2}$ couples to superconducting qubit with strength $g_{2}$.
        Moreover, we consider that the two magnon modes interact with the qubit in their respective cavities, and there is no interaction between the two magnon modes.
        The total Hamiltonian of our system can be written as ($\hbar = 1$) \cite{PhysRevB.100.134421}
        \begin{eqnarray}
            \label{eqn:h_0}
            \hat{H}_{0} & = & \omega_{q} \sigma^{\dagger} \sigma + \omega_{1} \hat{m}_{1}^{\dagger} \hat{m}_{1} + \omega_{2} \hat{m}_{2}^{\dagger} \hat{m}_{2} + g_{1} \left( \sigma^{\dagger} \hat{m}_{1} + \sigma \hat{m}_{1}^{\dagger} \right) \nonumber
            \\
            && + g_{2} \left( \sigma^{\dagger} \hat{m}_{2} + \sigma \hat{m}_{2}^{\dagger} \right) + \Omega \left( \hat{m}_{1}^{\dagger} e^{-i \omega_{d} t} + \hat{m}_{1} e^{i \omega_{d} t} \right).
        \end{eqnarray}
        where $\sigma$ is the ladder operator for the qubit.
        Here, the first term is the free Hamiltonian of the qubit, the second and third terms represent the energies of the two magnon modes and their corresponding interaction with the qubit is given by the fourth and fifth terms.
        The last term represents the driving energy of magnon mode $\hat{m}_{1}$ with driving strength $\Omega$.
        It may be noted that in deriving this Hamiltonian, we have eliminated the cavity modes via an adiabatic elimination (refer to Appendix \ref{app:adiabatic} for complete derivation).
        In a frame rotating at the driving frequency $\omega_{d}$, the Hamiltonian in Eq. \eqref{eqn:h_0} can be rewritten as
        \begin{eqnarray}
            \label{eqn:h}
            \hat{H} & = & \delta_{q} \sigma^{\dagger} \sigma + \delta_{1} \hat{m}_{1}^{\dagger} \hat{m}_{1} + \delta_{2} \hat{m}_{2}^{\dagger} \hat{m}_{2} + g_{1} \left( \sigma^{\dagger} \hat{m}_{1} + \sigma \hat{m}_{1}^{\dagger} \right) \nonumber
            \\
            && + g_{2} \left( \sigma^{\dagger} \hat{m}_{2} + \sigma \hat{m}_{2}^{\dagger} \right) + \Omega \left( \hat{m}_{1}^{\dagger} + \hat{m}_{1} \right),
        \end{eqnarray}
        where $\delta_{j} = \omega_{j} - \omega_{d}$ ($j \in \{ 1, 2 \}$ unless stated otherwise) are the detuning of the magnons in the two YIG sphere from the driving field and $\delta_{q} = \omega_{q} - \omega_{d}$ is the detuning of the qubit from the driving field.
       
        \subsection{Approximate analytical solution}
            \label{sec:model_analytical}
            \begin{figure}[ht]
                \centering
                \includegraphics[width=0.48\textwidth]{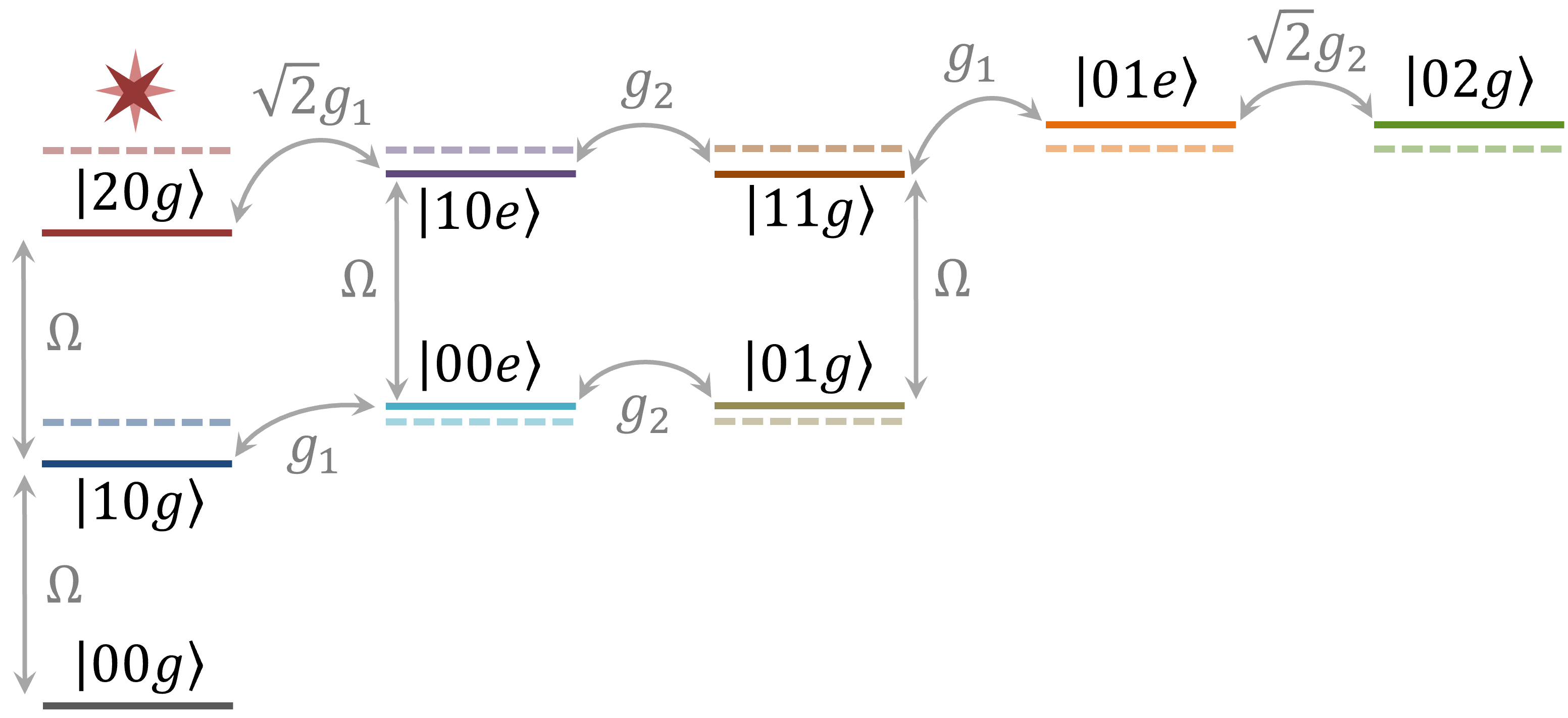}
                \caption{
                    Energy level diagram of the permitted states of the system under weak driving in the low-temperature limit.
                    Here, $g$ and $e$ denote the ground and excited states of the qubit.
                    The dashed lines denote the states at zero detunings.
                }
                \label{fig:energies}
            \end{figure}

            Under weak driving and in the low-temperature limit, the transitions permitted by the Hamiltonian in Eq. \eqref{eqn:h} results in the occupation of specific lower energy levels of the system as shown in Fig. \ref{fig:energies}.
            Thus, the truncated state of the system can be written as 
            \begin{eqnarray}
                \label{eqn:state}
                \ket{\Psi} & = & C_{0 0 g} \ket{0 0 g} + C_{1 0 g}\ket{1 0 g} + C_{0 0 e} \ket{0 0 e} \nonumber
                \\
                && + C_{0 1 g} \ket{0 1 g}+ C_{1 0 e} \ket{1 0 e} + C_{1 1 g}\ket{1 1 g} \nonumber
                \\
                && + C_{0 1 e} \ket{0 1 e} + C_{2 0 g} \ket{2 0 g} + C_{0 2 g}\ket{0 2 g},
            \end{eqnarray}
            where each $\ket{n_{1} n_{2} s}$ represents an energy state of the system ($n_{1}, n_{2} \in \{ 0, 1, 2 \}$ denote the number states of the magnon modes $\hat{m}_{1}$ and $\hat{m}_{2}$ respectively while $s \in \{ g, e \}$ denotes the qubit's ground or excited state).
            The coefficients $C_{n_{1} n_{2} s}$ are the corresponding probability amplitudes.
            To determine the time evolution of these coefficients under the effect of attenuation of the magnons (rate $\kappa_{1} = \kappa_{2} = \kappa$) and the qubit decay (rate $\gamma$), we consider the effective non-Hermitian Hamiltonian \cite{TaylorFrancis.QuantumOptomechanics.Bowen}
            \begin{equation}
                \label{eqn:h_eff}
                \hat{H}_{\mathrm{eff}} = \hat{H} - i \frac{\kappa}{2} \hat{m}_{1}^{\dagger} \hat{m}_{1} - i \frac{\kappa}{2} \hat{m}_{2}^{\dagger} \hat{m}_{2} - i \frac{\gamma}{2} \sigma^{\dagger} \sigma.
            \end{equation}
            and obtain the set of dynamic equations of the coefficients from the Schrodinger equation as
            \begin{subequations}
                \label{eqn:rates}
                \begin{eqnarray}
                    i \dot{C}_{0 0 g} & = & \Omega C_{1 0 g},
                    \\
                    i \dot{C}_{1 0 g} & = & \Delta_{1} C_{1 0 g} +g_{1} C_{0 0 e} + \Omega C_{0 0 g} + \sqrt{2} \Omega C_{2 0 g},
                    \\
                    i \dot{C}_{0 0 e} & = & \Delta_{q} C_{0 0 e} + g_{1} C_{1 0 g} + g_{2} C_{0 1 g} + \Omega C_{1 0 e},
                    \\
                    i \dot{C}_{0 1 g} & = & \Delta_{2} C_{0 1 g} + g_{2} C_{0 0 e} + \Omega C_{1 1 g},
                    \\
                    i \dot{C}_{2 0 g} & = & 2 \Delta_{1} C_{2 0 g} + \sqrt{2} g_{1} C_{1 0 e} + \sqrt{2} \Omega C_{1 0 g},
                    \\
                    i \dot{C}_{1 0 e} & = & \left( \Delta_{1} + \Delta_{q} \right) C_{1 0 e} + \sqrt{2} g_{1} C_{2 0 g} + g_{2} C_{1 1 g} \nonumber
                    \\
                    && + \Omega C_{0 0 e},
                    \\
                    i \dot{C}_{1 1 g} & = & \left( \Delta_{1} + \Delta_{2} \right) C_{1 1 g} + g_{2} C_{1 0 e} + g_{1} C_{0 1 e} \nonumber
                    \\
                    && + \Omega C_{0 1 g},
                    \\
                    i \dot{C}_{0 1 e} & = & \left( \Delta_{2} + \Delta_{q} \right) C_{0 1 e} + \sqrt{2} g_{2} C_{0 2 g} + g_{1} C_{1 1 g},
                    \\
                    i \dot{C}_{0 2 g} & = & 2 \Delta_{2} C_{0 2 g} + \sqrt{2} g_{2} C_{0 1 e},
                \end{eqnarray}
            \end{subequations}
            where $\Delta_{j} = \delta_{j} - \frac{i \kappa}{2}$ and $\Delta_{q} = \delta_{q} - \frac{i \gamma}{2}$.
            At this point, we assume that under weak driving, $C_{0 0 g} \gg \{ C_{1 0 g}, C_{0 0 e}, C_{0 1 g} \} \gg \{ C_{2 0 g}, C_{1 0 e}, C_{1 1 g}, C_{0 1 e}, C_{0 2 g} \}$.
            Thus, taking $C_{0 0 g} \approx 1$, we obtain the steady state solutions for the probability amplitudes and derive the zero time delay second-order correlation function for the magnon mode $\hat{m}_{1}$ as  
            \begin{eqnarray}
                \label{eqn:g_2_analytical}
                g^{(2)} (0) & = & 2 \frac{\left| C_{20g} \right|^{2}}{\left| C_{10g} \right|^{4}} \nonumber
                \\
                & = & \frac{2 \left| A_{0} B_{1} - A_{1} B_{0} \right|^{2} \left| A_{1} B_{2} - A_{2} B_{1} \right|^{2}}{\left| A_{2} B_{0} - A_{0} B_{2} \right|^{4}},
            \end{eqnarray}
            where $A_{i}$ and $B_{i}$ ($i \in \{ 0, 1, 2 \}$) are the coefficients of the simplified set of linear equations in $C_{10g}$ and $C_{20g}$ (refer to Appendix \ref{app:optimal} for detailed expressions).

            Here, $g^{(2)} (0)$ gives the probability of finding two magnons in the magnon mode $\hat{m}_{1}$ at the same time.
            Magnon antibunching can thereby be obtained when $g^{(2)} (0) < 1$ with complete magnon blockade when $g^{(2)} (0) \to 0$.
            Physically, when $g^{(2)} (0) < 1$, the magnons exhibit a sub-Poissonian distribution as the excitation of the first magnon prevents the simultaneous excitation or transmission of a second magnon in the system, which results in an orderly output of magnons one after another.
            This phenomenon is a result of the destructive quantum interference between the excitation pathways (i) $\ket{00g} \to \ket{10g} \to \ket{20g}$, (ii) $\ket{00g} \to \ket{10g} \to \ket{00e} \to \ket{10e} \to \ket{20g}$ and (iii) $\ket{00g} \to \ket{10g}\to \ket{00e} \to \ket{01g} \to \ket{11g} \to \ket{10e} \to \ket{20g}$.
            The resultant two-magnon state assumes a very low probability amplitude, resulting in the observation of magnon blockade.
            An explicit condition can therefore be derived from Eq. \eqref{eqn:optimal} as (refer to Appendix \ref{app:optimal} for derivation)
            \begin{eqnarray}
                \label{eqn:optimal}
                \Delta_{2}\tilde{\Delta}_{2}( \Delta_{2}\tilde{\Delta}_{2} \Delta_{2}^{\prime} - g_{1}^{2} \tilde{\Delta}_{2} -g_{1}^{2}\Delta_{q})\nonumber
                \\
                + (\Delta_{1}\Delta_{2}\tilde{\Delta}_{2} - \Omega^{2}\Delta_{s} + g_{1}^{2} \Delta_{2})(\Delta_{2}^{\prime} \Delta_{s} - g_{1}^{2})=0
            \end{eqnarray}
        
       \subsection{Exact numerical solution}
            \label{sec:model_numerical}
            The steady-state zero time delay second-order correlation function  can also be obtained numerically by solving the quantum Lindblad master equation \cite{carmichael1999dissipation}
            \begin{eqnarray}
                \label{eqn:master}
                \dot{\rho} = - i \left[ \hat{H}, \rho \right] + \frac{\kappa}{2} \left( n_{th_{1}}+1 \right) \mathcal{L} \left[ \hat{m}_{1} \right] \rho + \frac{\kappa}{2} \left( n_{th_{1}} \right) \mathcal{L} \left[ \hat{m}_{1}^{\dagger} \right] \rho \nonumber
                \\
                +\frac{\kappa}{2} \left( n_{th_{2}} + 1 \right) \mathcal{L} \left[ \hat{m}_{2} \right] \rho + \frac{\kappa}{2} \left( n_{th_{2}} \right) \mathcal{L} \left[ \hat{m}_{2}^{\dagger} \right] \rho + 
                \frac{\gamma}{2} \mathcal{L} \left[ \sigma \right] \rho,
            \end{eqnarray}
            where $\mathcal{L} [ \hat{\mathcal{O}} ] \rho = 2 \hat{\mathcal{O}}\rho \hat{\mathcal{O}}^{\dagger} - \hat{\mathcal{O}}^{\dagger} \hat{\mathcal{O}} \rho - \rho \hat{\mathcal{O}}^{\dagger} \hat{\mathcal{O}}$ represents the Lindblad operator ($\hat{\mathcal{O}} \in \{ \hat{m}_{1}, \hat{m}_{2}, \sigma \}$).
            Here, $n_{\mathrm{th}_{j}}= \{ \exp{( \frac{\hbar \omega_{j}}{k_{\mathrm{B}} T} )} -1 \}^{-1}$ represent the mean thermal magnon number at temperature $T$.
            From Eq. \eqref{eqn:master}, one can obtain \cite{JOSAB.33.1322}
            \begin{equation}
                \label{eqn:g_2_numerical}
                g^{(2)} (0) = \frac{\mathrm{Tr} \left[ \rho \hat{m}_{1}^{\dagger} \hat{m}_{1}^{\dagger} \hat{m}_{1} \hat{m}_{1} \right]}{\left( \mathrm{Tr} \left[ \rho \hat{m}_{1}^{\dagger} \hat{m}_{1} \right] \right)^{2}}.
            \end{equation}

    \section{Results}
        \label{sec:results}
        \begin{figure}[ht]
            \centering
            \includegraphics[width=0.48\textwidth]{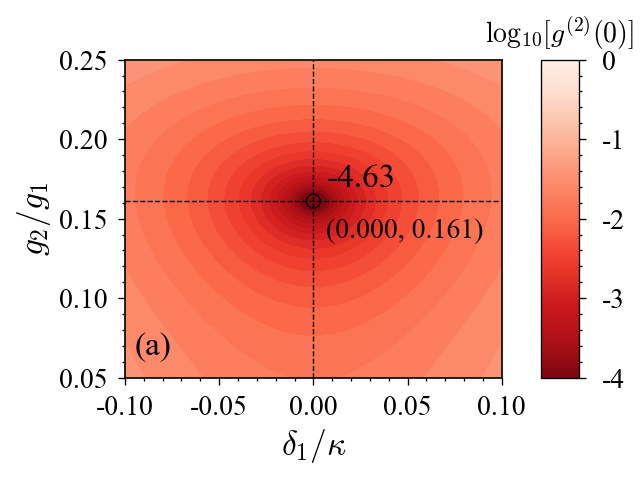}
            \includegraphics[width=0.48\textwidth]{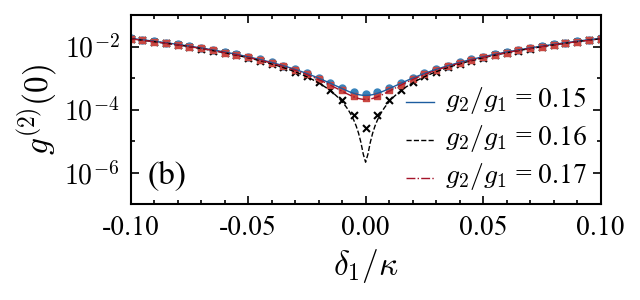}
            \includegraphics[width=0.48\textwidth]{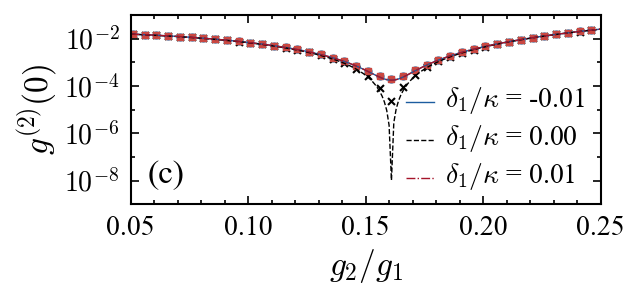}
            \caption{
                (Color online) (a) Zero time delay second-order correlation $g^{(2)} (0)$ with variation in the the coupling strength ratio $g_{2} / g_{1}$ and the magnon detuning $\delta_{1}$ obtained using Eq. \eqref{eqn:g_2_numerical} when $\delta_{q} = \delta_{2} = 0$.
                The dashed lines denote the minimas for fixed values of $\delta_{1}$ and $g_{2} / g_{1}$.
                (b) Behaviour of $g^{(2)} (0)$ with variation in $\delta_{1}$ for different values of coupling strength ratio $g_{2} / g_{1}$.
                (c) Behaviour of $g^{(2)} (0)$ with variation in $g_{2} / g_{1}$ for different values of $\delta_{1}$.
                The solid lines denote the values obtained using the approximate expression in Eq. \eqref{eqn:g_2_analytical}.
                The corresponding scatter points represent the respective values obtained from (a).
                Other parameters (in units of $\kappa$) are $\gamma = 1.11$, $g_{1} = 0.8$ and $\Omega = 0.001$.
            }
            \label{fig:g_2_0s_0}
        \end{figure}

        \begin{figure}[ht]
            \centering
            \includegraphics[width=0.48\textwidth]{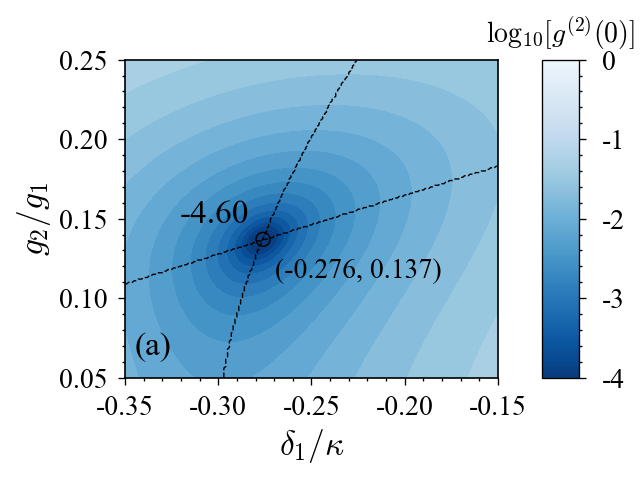}
            \includegraphics[width=0.48\textwidth]{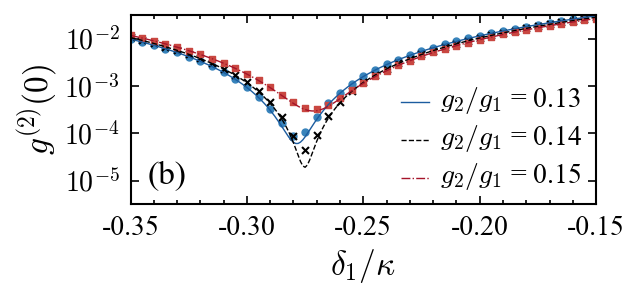}
            \includegraphics[width=0.48\textwidth]{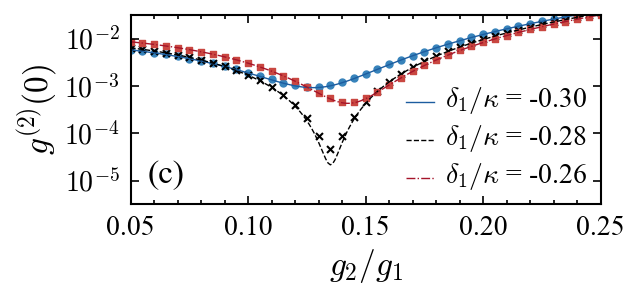}
            \caption{
                (Color online) (a) Numerical values of zero time delay second-order correlation $g^{(2)} (0)$ with variation in the magnon detuning $\delta_{1}$ and the coupling strength ratio $g_{2} / g_{1}$ when $\delta_{q} = \delta_{2} = 0.1\kappa$.
                The dashed lines denote the minimas for fixed values of $\delta_{1}$ and $g_{2} / g_{1}$.
                (b) Plot of $g^{(2)} (0)$ obtained analytically (solid lines) and numerically (scatter plots) with $\delta_{1}$ for different values of $g_{2} / g_{1}$.
                (c) Variation of $g^{(2)} (0)$ with respect to $g_{2} / g_{1}$ for different values of $\delta_{1}$.
                Other parameters are the same as Fig. \ref{fig:g_2_0s_0}.
            }
            \label{fig:g_2_0s_1}
        \end{figure}

        We obtain the exact numerical values of Eq. \eqref{eqn:g_2_numerical} using QuTiP's steady-state solvers for the master equation (Eq. \eqref{eqn:master}) \cite{ComputPhysCommun.183.1760, ComputPhysCommun.184.1234} and compare it with the analytically obtained values from Eq. \eqref{eqn:g_2_analytical}.
        For our simulations, we consider a weak driving strength of $\Omega = 0.001 \kappa$, with the magnon decay rate $\kappa = 2 \pi \times 1.8$ MHz and the qubit decay rate $\gamma = 1.11 \kappa$ \cite{CRPhys.17.729}.
        We also fix the coupling between the driven magnon and the qubit at $g_{1} = 0.8 \kappa$ \cite{JOSAB.38.876}.
        With these parameters, we analyze two different scenarios for the magnon antibunching in mode $\hat{m}_{1}$.
        In the first scenario, the undriven magnon and the qubit are zero detuned from the driving optical field. i.e., $\delta_{2} = 0$ and $\delta_{q} = 0$.
        In the second scenario, the undriven magnon and the qubit have non-zero finite detunings.

        \subsection{Antibunching at zero detunings}
            \label{sec:results_zero}
            When the undriven magnon and the qubit are in resonance with the driving field ($\delta_{2} = \delta_{q} = 0$), we observe that for any fixed value of the magnon-qubit coupling strength ratio, $g_{2}/g_{1}$, maximum antibunching is achieved when the detuning of the driven magnon is also in resonance with the driving field, i.e., $\delta_{1} / \kappa = 0$.
            Moreover, for our choice of parameters, we find that $g^{(2)} (0)$ is minimum when $g_{2} = 0.161 g_{1}$.
            We illustrate this behavior obtained from Eq. \eqref{eqn:g_2_numerical} in Fig. \ref{fig:g_2_0s_0} (a), where the dashed lines represent the optimal values of $\delta_{1}$ and $g_{2} / g_{1}$.
            In Fig. \ref{fig:g_2_0s_0} (b) and \ref{fig:g_2_0s_0} (c), we compare this observation with the results of our approximate analytical expression.
            It may be noted here that the minimum values of $g^{(2)} (0)$ obtained from the analytical expression differs from those obtained from the numerical simulations.
            This may be attributed to the fact that while deriving Eq. \eqref{eqn:g_2_analytical}, we have assumed a higher occupancy of the ground state than the excited states, with $C_{00g} \approx 1$.
            This assumption results in fewer excitations to the $\ket{20g}$ state, which leads to lower values of $g^{(2)} (0)$.
            Nonetheless, the location of these minimums is in agreement with either approach.
            We utilize this characteristic feature in later sections to compare the optimal positions of antibunching for different sets of tunable parameters.

        \subsection{Antibunching at finite detunings}
            \label{sec:results_finite}
            In Fig. \ref{fig:g_2_0s_1} (a), we plot $g^{(2)} (0)$ at finite magnon and qubit detunings of $\delta_{2} = 0.1 \kappa$ and $\delta_{q} = 0.1 \kappa$ respectively.
            In this case, we observe a global minima at $\delta_{1} = -0.276 \kappa$ and $g_{2} = 0.137 g_{1}$.
            The behavior of antibunching around this minima for fixed values of $g_{2}/g_{1}$ and $\delta_{1}/\kappa$ is illustrated in Fig. \ref{fig:g_2_0s_1} (b) and \ref{fig:g_2_0s_1} (c) respectively.
            Unlike the situation at resonance, here we observe that for higher values of $g_{2} / g_{1}$, the optimal detuning of the magnon mode $\hat{m}_{1}$ shifts monotonically towards lower blue-detuned values.
            This ``shift'' may be explained in terms of quantum interference as follows.
            As discussed in Section \ref{sec:model_analytical}, magnon antibunching in our system can be primarily attributed to the destructive interference between three distinct pathways.
            Using the probability amplitudes, it can be analytically shown that the pathway $\ket{00g} \to \ket{10g} \to \ket{00e} \to \ket{10e} \to \ket{20g}$ interferes destructively whereas the pathway $\ket{00g} \to \ket{10g} \to \ket{00e} \to \ket{01g} \to \ket{11g} \to \ket{10e} \to \ket{20g}$ interferes constructively with the primary pathway $\ket{00g} \to \ket{10g} \to \ket{20g}$.
            Moreover, the probability of one of these pathways, namely $\ket{00g} \to \ket{10g} \to \ket{00e} \to \ket{01g} \to \ket{11g} \to \ket{10e} \to \ket{20g}$, depends largely on the magnitude of the coupling constant $g_{2}$.
            Higher values of the coupling $g_{2}$ result in higher probabilities of the states $\ket{01g}$ and $\ket{11g}$, as can be seen from Fig. \ref{fig:energies}.
            It can also be seen from Fig. \ref{fig:energies} that the detuning $\delta_{1}$ mainly effects the difference between the energy levels $\ket{10g}$ and $\ket{00g}$ (or $\ket{10e}$ and $\ket{20g}$).
            For lower blue-detuned values, the energy levels move closer to one another, enhancing their corresponding transition pathways.
            The constructive interference enhanced due to an increase in the coupling strength $g_{2}$ is therefore compensated with the destructive interference enhanced by shifting $\delta_{1}$ to lower blue-detuned values.
            Thus, by tuning the coupling strength $g_{2}$ and driving detuning $\delta_{1}$, one can regulate the degree of antibunching and realize magnon blockade in this scenario.
            
\begin{figure}[ht]
                \centering
                \includegraphics[width=0.48\textwidth]{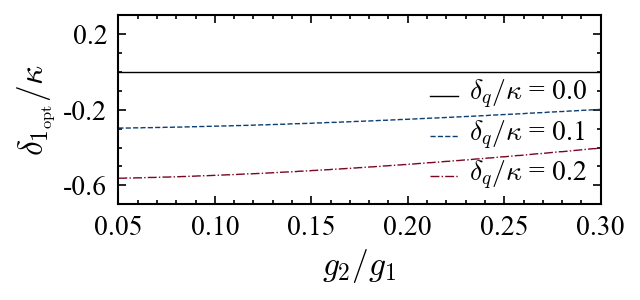}
                \caption{
                    (Color online) Magnon detuning $\delta_{1}$ at which the highest antibunching is observed for different values of the coupling strength ratio $g_{2} / g_{1}$ and the qubit detuning $\delta_{q}$ obtained using Eq. \eqref{eqn:optimal}.
                    Here, $\delta_{2} = \delta_{q}$ with other parameters same as Fig. \ref{fig:g_2_0s_1}.
                }
                \label{fig:optimal}
            \end{figure}

\begin{figure}[ht]
                \centering
                \includegraphics[width=0.48\textwidth]{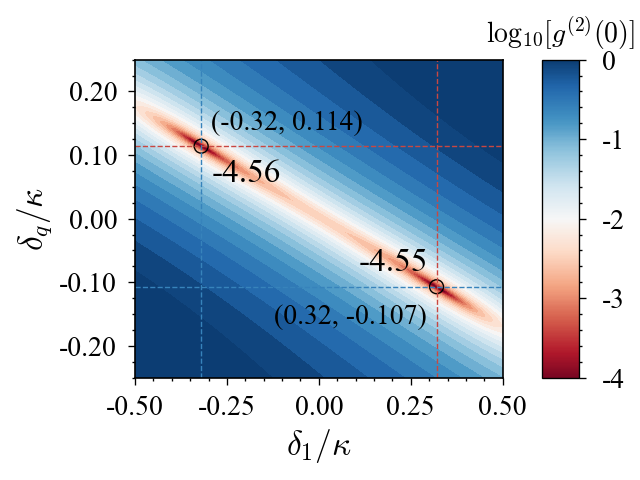}
                \caption{
                    (Color online) Numerical values of the zero time delay second-order correlation $g^{(2)} (0)$ with variation in the qubit detuning $\delta_{q}$ and the driven magnon detuning $\delta_{1}$.
                    Here, $\delta_{2} = 0.1 \kappa$ and $g_{2} = 0.125 g_{1}$.
                    Other parameters are same as Fig. \ref{fig:g_2_0s_0}.
                }
                \label{fig:g_2_0s_deltas}
            \end{figure}

            It may also be noted that the optimal value of detuning $\delta_{1}$ shifts closer or away from the resonance depending on the value of $\delta_{q}$ and $\delta_{2}$.
            For a qualitative insight on this variation, we plot the optimal detuning $\delta_{1_{\mathrm{opt}}}$ against the coupling strength ratio $g_{2}/g_{1}$ for different values of $\delta_{q}$ ($= \delta_{2}$) in Fig. \ref{fig:optimal}.
            It can be seen here that for higher red-detuned values of qubit (and undriven magnon) detuning, maximum antibunching is achieved at higher blue-detuned values of $\delta_{1_{\mathrm{opt}}}$.
            As $\delta_{q}, \delta_{2} \rightarrow 0$, this optimal value also approaches zero, which agrees with the above-mentioned resonant scenario.
            In addition to this, we observe that for a fixed configuration of the undriven YIG sphere, i.e., for fixed values of $\delta_{2}$ and $g_{2}$, the location of magnon blockade depends on the type of detuning of the qubit and the driven magnon.
            More specifically, when the qubit is red-detuned (blue-detuned), the optimal location for the detuning of the driven magnon lies in the blue-sideband (red-sideband), as shown in Fig. \ref{fig:g_2_0s_deltas}.
        
            \begin{figure}[ht]
                \centering
                \includegraphics[width=0.48\textwidth]{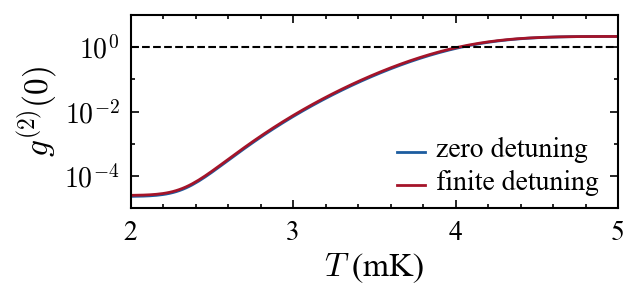}
                \caption{
                    (Color online) Plot of zero time delay second-order correlation $g^{(2)} (0)$ with variation in mean thermal magnon number for zero detunings (red) and for finite detunings (blue) at the optimal values marked in Fig. \ref{fig:g_2_0s_0} (a) and Fig. \ref{fig:g_2_0s_1} (a) respectively.
                    The dashed black line represents $g^{(2)} (0) = 1$.
                    Other parameters are same as Fig. \ref{fig:g_2_0s_0}.}
                \label{fig:thermal}
            \end{figure}
            
            Thermal magnon population has a considerable effect on the magnon blockade characteristics.
            Therefore, we examine equal time second order correlation $g^{(2)}(0)$ by taking into account the mean thermal magnon occupation number $n_{\mathrm{th}_{1}}$ and $n_{\mathrm{th}_{2}}$ in the exact numerical simulation of the quantum master equation.
            We choose the magnon frequencies as $\omega_{1} / 2 \pi = 8.2$ GHz and $\omega_{2} / 2 \pi = 8.6$ GHz \cite{CRPhys.17.729,PhysRevB.100.134402}. 
            In Fig. \ref{fig:thermal}, we plot the variation of $g^{(2)}(0)$ with temperature $T$ for both the cases of zero detunings and finite detunings.
            We observe that for either case, $g^{(2)} (0) < 1$ for upto $T \approx 4$ mK.
            Therefore, the YIG spheres in the system must be maintained at low temperatures to enhance magnon antibunching and observe magnon blockade.

    \section{Conclusion}
        \label{sec:conclusion}
        In this work, we theoretically studied quantum interference-induced magnon blockade in a system containing two YIG spheres and a superconducting qubit inside two mutually perpendicular microwave cavities. The magnon modes in the two YIG spheres were indirectly coupled to the superconducting qubit via virtual photon excitations of the cavity modes.
        We analyzed the zero time delay second-order correlations in a driven magnon mode by varying detunings and coupling strengths. We discussed the blockade and antibunching characteristics of the driven magnon (i) at resonance and (ii) at finite detunings.
        In the weak coupling regime, we showed that the magnon antibunching in the driven mode could be tuned by adjusting the coupling strength between the undriven magnon mode and the qubit.
        The detunings of the magnons and the qubit play an important role in the degree of achievable antibunching and magnon blockade.
        We find that the optimal detuning of driven magnon depends on the detunings of undriven magnon and the qubit.
        Finally we discuss the effect of thermal noise on magnon blockade and antibunching characteristics.
        Magnon blockade is an efficient way to realise single magnons, which have several applications in quantum information processing and the preparation of non-classical states \cite{ApplPhysLett.122.084001}.
        We believe our work will contribute towards the experimental realization of single magnon sources for emerging quantum technologies.

    \section*{Acknowledgment}
        S.K. would like to acknowledge MHRD, Government of India for providing financial support for the research through PMRF scheme.
        
    \appendix
    \section{Adiabatic elimination}
        \label{app:adiabatic}
        The two magnon modes and the superconducting qubit interact directly with the microwave cavity modes respectively via magnetic dipole and electric dipole interactions.
        The complete Hamiltonian in the rotating frame at driving frequency $\omega_{d}$ can be written as
        \begin{eqnarray}
            \label{eqn:h_f}
            \hat{H}_{f} & = & \Sigma_{j} \Big\{ \delta_{0j} \hat{a}_{j}^{\dagger} \hat{a}_{j} + \delta_{j}^{\prime} \hat{m}_{j}^{\dagger} \hat{m}_{j} + g_{m_{j}} \left( \hat{m}_{j}^{\dagger} \hat{a}_{j} + \hat{m}_{j} \hat{a}_{j}^{\dagger} \right) \nonumber
            \\
            && + g_{q_{j}} \left( \sigma^{\dagger} \hat{a}_{j} + \sigma \hat{a}_{j}^{\dagger} \right) \Big\} + \delta_{q}^{\prime} \sigma^{\dagger} \sigma \nonumber
            \\
            && + \Omega \left( \hat{m}_{1}^{\dagger} + \hat{m}_{1} \right)
        \end{eqnarray}

        where $\delta_{0j}$ ($\delta_{j}^{\prime}$) are the corresponding detunings of the microwave cavity (magnon) modes $\hat{a}_{j}$ ($\hat{m}_{j}$) from the drive frequency and $\delta_{q}^{\prime}$ is the detuning of superconducting qubit.
        $g_{m_{j}}$ denote the coupling strengths between the magnon and cavity modes and $g_{q_{j}}$ denote the coupling strengths between the qubit and cavity modes.
        $\Omega$ is the driving amplitude.

        The dynamics of the cavity modes $\hat{a}_{j}$ are therefore governed by the quantum Langevin equation \cite{TaylorFrancis.QuantumOptomechanics.Bowen}
        \begin{eqnarray}
            \dot{\hat{a}}_{j} & = & - i \delta_{0j} \hat{a}_{j} -i g_{q_{j}} \sigma - i g_{m_{j}} \hat{m}_{j}
        \end{eqnarray}

        In the limit where $\delta_{0j} \gg \delta_{q}, \delta_{j}^{\prime}, g_{q_{j}}, g_{m_{j}}, \Omega$, we can adiabatically eliminate the cavity modes $\hat{a}_{j}$ and set $\dot{\hat{a}}_{j} \approx 0$ \cite{PhysRevA.71.012331}.
        The steady-state values of the cavity modes can therefore be written as
        \begin{eqnarray}
            \hat{a}_{j} & = & - \frac{g_{q_{j}} \sigma}{\delta_{0j}} - \frac{g_{m_{j}} \hat{m}_{j1}}{\delta_{0j}}
        \end{eqnarray}
        
        Substituting these values in the Hamiltonian of Eq. \eqref{eqn:h_f}, we obtain

        \begin{eqnarray}
            \label{eqn:h_ad}
            \hat{H} & = & \delta_{q} \sigma^{\dagger} \sigma + \delta_{1} \hat{m}_{1}^{\dagger} \hat{m}_{1} + \delta_{2} \hat{m}_{2}^{\dagger} \hat{m}_{2} + g_{1} \left( \sigma^{\dagger} \hat{m}_{1} + \sigma \hat{m}_{1}^{\dagger} \right) \nonumber
            \\
            && + g_{2} \left( \sigma^{\dagger} \hat{m}_{2} + \sigma \hat{m}_{2}^{\dagger} \right) + \Omega \left( \hat{m}_{1}^{\dagger} + \hat{m}_{1} \right),
        \end{eqnarray}
        with
        \begin{subequations}
            \begin{eqnarray}
                \delta_{q} & = & \delta_{q}^{\prime} - \frac{g_{q_{1}}^{2}}{\delta_{01}} - \frac{g_{q_{2}}^{2}}{\delta_{02}}
                \\
                \delta_{j} & = & \delta_{j}^{\prime} -\frac{g_{m_{j}}^{2}}{\delta_{0j}}
                \\
                g_{j} & = & - \frac{g_{m_{j}} g_{q_{j}}}{\delta_{0j}}
             \end{eqnarray}
        \end{subequations}

        In the Hamiltonian of Eq. \ref{eqn:h_ad}, the first three terms represent the free energy of qubit and the two magnon modes, the fourth and fifth terms denote the indirect interactions of the qubit and magnon modes through virtual photon excitation with effective coupling strength $g_{j}$ and the last term represents the driving energy.
           
    \section{Derivation of optimal equation for magnon blockade}
        \label{app:optimal}
        From Eq. \eqref{eqn:rates}, we obtain a simplified set of linear equations $A_{0} + A_{1} C_{10g} + A_{2} C_{20g} = 0$ and $B_{0} + B_{1} C_{10g} + B_{2} C_{20g} = 0$, where the coefficients $A_{i}$'s and $B_{i}$'s are
        \begin{subequations}
            \begin{eqnarray}
                A_{0} & = & \Omega \Delta_{2} \tilde{\Delta}_{2}
                \\
                A_{1} & = & \Delta_{1} \Delta_{2} \tilde{\Delta}_{2} + \Omega^{2} \Delta_{s} - g_{1}^{2} \Delta_{2}
                \\
                A_{2} & = & \sqrt{2} \Omega \left( \Delta_{1} \Delta_{s} + \Delta_{2} \tilde{\Delta}_{2} - g_{1}^{2} \right)
                \\
                B_{0} & = & \Omega^{2} \left( \Delta_{2}^{\prime} \Delta_{s} - g_{1}^{2} \right)
                \\
                B_{1} & = & \Omega \Big\{ 2 \Delta_{1} \left( \Delta_{2}^{\prime} \Delta_{s}-g_{1}^{2} \right) \nonumber
                \\
                && + \Delta_{2}^{\prime} \left( \Delta_{2} \tilde{\Delta}_{2} - g_{1}^{2} \right) - g_{1}^{2} \Delta_{q} \Big\} 
                \\
                B_{2} & = & \sqrt{2} \left( \Delta_{1}^{2} + \Delta_{1} \tilde{\Delta}_{1} \right) \left\{ \Delta_{2}^{\prime} \left( \Delta_{1} + \Delta_{2} \right) - g_{1}^{2} \right\} \nonumber
                \\
                && + \sqrt{2} \Omega^{2} \Delta_{2}^{\prime} \left( \Delta_{1} + \Delta_{q} \right) - \sqrt{2} g_{2}^{2} \Delta_{1} \Delta_{2}^{\prime}
            \end{eqnarray}
        \end{subequations}
        where we have used the substitutions
        \begin{subequations}
            \begin{eqnarray}
               \Delta_{s} & = & \Delta_{1} + \Delta_{2} + \Delta_{q} 
               \\
               \Delta_{j}^{\prime} & = & \Delta_{j} + \Delta_{q} - \frac{g_{j}^{2}}{\Delta_{j}}
               \\
               \tilde{\Delta}{j} & = & \Delta_{q} + \frac{\Omega^{2}}{\Delta_{j}} - \frac{g_{j}^{2}}{\Delta_{j}}
            \end{eqnarray}
        \end{subequations}

        Solving the linear equations gives us the coefficients
        \begin{subequations}
            \begin{eqnarray}
                C_{10g} & = & \frac{A_{2} B_{0} - A_{0} B_{2}}{A_{1} B_{2} - A_{2} B_{1}}
                \\
                C_{20g} & = & \frac{A_{0} B_{1} - A_{1} B_{0}}{A_{1} B_{2} - A_{2} B_{1}}
            \end{eqnarray}
        \end{subequations}
        where we have assumed $A_{2} \neq 0$ and $| A_{1} B_{2} - A_{2} B_{1} |^{2} \neq 0$.
        Substituting these values in $g^{(2)} (0) = 2 | C_{20g} |^{2} / | C_{10g} |^{4}$, we obtain Eq. \eqref{eqn:g_2_analytical}, from which we get the condition for blockade as
        \begin{equation}
            \label{eqn:optimal_alt}
                \left| A_{0} B_{1} - A_{1} B_{0} \right|^{2} = 0 
        \end{equation}
        or equivalently, Eq. \eqref{eqn:optimal}.

    \bibliography{manuscript}

\end{document}